# Tracking single particles on supported lipid membranes: multi-mobility diffusion and nanoscopic confinement

Chia-Lung Hsieh*, Susann Spindler, Jens Ehrig, Vahid Sandoghdar

*Max Planck Institute for the Science of Light and Friedrich Alexander University, 91058 Erlangen, Germany*
*\* Current address: Institute of Atomic and Molecular Sciences, Academia Sinica, Taipei 106, Taiwan*

**Abstract**

Supported lipid bilayers have been studied intensively over the past two decades. In this work, we study the diffusion of single gold nanoparticles (GNPs) with diameter of 20 nm attached to GM1 ganglioside or DOPE lipids at different concentrations in supported DOPC bilayers. The indefinite photostability of GNPs combined with the high sensitivity of interferometric scattering microscopy (iSCAT) allows us to achieve 1.9 nm spatial precision at 1 ms temporal resolution, while maintaining long recording times. Our trajectories visualize strong transient confinements within domains as small as 20 nm, and the statistical analysis of the data reveals multiple mobilities and deviations from normal diffusion. We present a detailed analysis of our findings and provide interpretations regarding the effect of the supporting substrate and GM1 clustering. We also comment on the use of high-speed iSCAT for investigating diffusion of lipids, proteins or viruses in lipid membranes with unprecedented spatial and temporal resolution.

**INTRODUCTION**

Membrane processes play a crucial role in the functionality of biological cells.[1-3] In addition, lipid bilayers provide a convenient configuration for studying fundamental diffusion and transport phenomena in two dimensions. While in an ideal system one would expect normal diffusion, barriers, traps or heterogeneities in realistic arrangements would result in anomalies.[4,5] The robust identification of deviations from normal diffusion depends, however, on the time and length scales of the underlying interactions. In the celebrated case of "lipid rafts", for example, one still struggles to resolve local nanoscopic confinements in real space and time.[6] Next to biological membranes, supported lipid bilayers (SLBs) have attracted the attention of scientists because they provide simple model systems for controlled studies. Interestingly, interactions with the underlying substrate have been reported to cause deviations from free diffusion also in this case.[7] Thus, quantitative measurements of the mobility of lipids and proteins, especially at high temporal and spatial resolution, are in general necessary for better understanding of the membrane structure and function.[8]

One of the most powerful tools for investigating diffusion in membranes is to trace the motion of a single particle attached to the lipids or proteins of interest. This technique, known as single-particle tracking (SPT), can measure the dynamics of phenomena such as thermal fluctuations, fluidity and organization of membranes without ensemble averaging.[9] Below, we provide a concise review of the recent developments in SPT. In particular, we point out the limitations of photostability, saturation and signal-to-noise ratio (SNR) on SPT as well as the challenges in the combination of spatial and temporal resolutions. We will show that interferometric detection of scattering (iSCAT) from very small nanoparticles can offer the ideal compromise. Using this method, we visualize nanoscopic confinement in SLBs and analyze deviations of the trajectory from the expectations of normal diffusion.

**SINGLE-PARTICLE TRACKING**

In SPT, optical labels are commonly used to distinguish the molecule to be studied from the wealth of other biological entities in the system. Different contrast mechanisms that are typically employed include fluorescence,[10-16] scattering,[17-20] and absorption.[21,22] The position of the optical label is determined by localizing the center of its point-spread function, whereby the precision depends on the SNR. With sufficient SNR, the object can be localized with precision much better than the diffraction limit of an optical microscope. In the ideal situation, the SNR is limited by the shot noise in the optical signal. In addition to a high precision in determining the particle position, diffusion studies also need a high temporal resolution. However, this requirement is intrinsically at odds with high SNR because smaller integration times lead to lower signal and SNR.

Early SPT efforts used fluorescent bead labels of several hundred nanometers in size.[15] Such small particles are still very large on the scale of individual lipids and proteins, and the substantial interface area with the membrane may lead to multiple bonds, drag forces or geometric hindrance of biochemical activities. Hence, it is desirable to use smaller particles or single dye molecules. However, as the number of fluorophores is reduced, the signal drops dramatically (proportional to the bead volume). In the ultimate case of a single molecule, saturation puts a limit on the number of photons that constitute the signal per unit time. Furthermore, photobleaching restricts the total number of photons that are emitted before the label becomes silent. As a result, one can choose to measure for a very short time at a high SNR and thus good localization precision per unit time or spread the photons over a longer time at the cost of a lower localization precision per unit time. Tero *et al.* chose the latter strategy and successfully tracked individual dye-labeled lipids in model membranes with 21 nm spatial precision at 2 kHz.[14] In this study, the total number of steps for individual trajectories was below 400, corresponding to an overall measurement time of 200 ms.[14]

Quantum dots (Qdots) are alternative fluorescent labels with significantly higher resistance to photobleaching. Compared to dye molecules, Qdots are larger in size, typically 15 to 25 nm in diameter including the surface functionalization. Using Qdot labeling, Clausen *et al.* tracked plasma membrane molecules at 1.75 kHz with 30 nm precision for up to 2 seconds (3500 steps).[13] To get around the limitations of fluorescence-based SPT, scientists have explored scattering as a contrast mechanism,[23] which offers two decisive advantages. First, lack of saturation in scattering (scattering signal is linearly proportional to incident intensity) makes it possible to increase the signal by raising the illuminating optical power, thus allowing for short integration times. Second, because the scattering signal is stable without photobleaching effect, this approach can handle unlimited observation times.

Metallic nanoparticles have been employed as labels due to their enhanced scattering at the surface plasmon resonance.[24] Using bright-field microscopy, Fujiwara *et al.* tracked 40 nm gold nanoparticles (GNPs) at 40 kHz with 17 nm precision on a cell plasma membrane.[18] At such a spatiotemporal resolution, plasma membranes were found to be compartmentalized into 230 nm zones by the actin-based membrane skeleton meshwork underneath.[18] Using dark-field microscopy, Ueno *et al.* were able to track 40 nm GNPs attached to motor proteins at < 2nm precision at 110 kHz.[19] A challenge in scattering-based SPT, however, is that the scattering signal drops as the sixth power of the particle size, putting a lower limit on the particle size that can be used. Reduction of the particle diameter by ten times would come at the impressive cost of $10^6$ fold loss in the signal.[25] Such a weak signal can be overwhelmed by the read-out noise of high-speed devices. As a result, particles smaller than 40 nm in diameter have rarely been used in scattering-based high-speed localization although studies have observed different behavior when using larger GNPs.[26] To address this issue, scientists have applied photothermal microscopy[21,22] to detect small GNPs even below 5 nm. However, the imaging speed of this technique is rather low (reported 11 nm precision at 25 Hz in Ref. [22]) and inherently limited since it uses a scanning imaging scheme. Furthermore, local heating restricts the applicability of this method.

## iSCAT MICROSCOPY

In our laboratory, we have developed iSCAT microscopy as a powerful approach for high-speed SPT.[25,27-29] This method keeps all the benefits of scattering-based detection, including the photostable signal and the unlimited observation time. In addition, taking advantage of the coherent nature of the scattering signal, iSCAT microscopy detects weak scattering signals from very small particles through a homodyne measurement based on the interference of a weak signal with a strong local oscillator. The interferometric detection pushes the system performance into the shot-noise-limited regime, where the SNR becomes proportional to $\sqrt{N}$ for $N$ detected photons per unit time. As a result, the electronic noise becomes negligible for large enough signals. The high SNR of iSCAT gives access to high localization precision of the point-spread function and high temporal resolution.

The iSCAT microscope (see Figure **1**A) is a home-built inverted optical microscope using a continuous-wave laser at 532 nm wavelength (Verdi G2, Coherent, Santa Clara, CA) for illumination. The laser light is focused at the back focal plane of an oil-immersion objective (UPLSAPO 100XO, NA1.4, Olympus, Tokyo, Japan), thereby creating a wide-field illumination on the sample over an area of ~ 5 µm in diameter. To increase the area of flat illumination, the laser beam is scanned by 4 mrad in both lateral directions using two-axis acousto-optic deflectors (DTSXY-400, AA Opto-Electronic, Orsay, France). The two deflectors scan the beam at ~100 kHz at slightly different frequencies, controlled by two synchronized function generators that are triggered by the CMOS camera (MV-D1024E-CL, Photonfocus AG, Lachen, Switzerland). Scanning the beam yields a homogeneous illumination over a field of view of 10×10 µm$^2$ (with a slow intensity decrease of < 10% at the edge). Images are recorded at 1000 frames per second with exposure time of 0.56 ms. The typical excitation intensity at the sample is 8 kW/cm$^2$.

The laser beam illuminates the sample consisting of a lipid bilayer deposited on a cover glass. The GNP bound to the lipid membrane resonantly scatters the incident light, which can be detected in reflection. The partial reflection at the water-glass interface due to the refractive index mismatch serves as the reference beam in a homodyne detection scheme. The backscattered light and the reflected reference beam are both collected by the objective, reflected by a beam-splitter, and imaged on the CMOS camera which records their interference signal. The heating of a 20 nm gold nanoparticle in water under illumination is negligible, as the temperature raised on the surface of the particle is estimated to be only 0.2 K for our experimental parameters.[30]

In the present work, we demonstrate iSCAT high-speed SPT for studying diffusion of guest lipid molecules in DOPC SLBs. The main system of choice is that of GM1 ganglioside, which has been of great interest in biological membranes, partly because of its close relationship with lipid raft domains in cell membranes.[31,32] Various techniques have been used to study how GM1 diffuses and organizes in membranes, including fluorescence microscopy,[33,34] SPT,[35,36] atomic force microscopy (AFM),[34,37-39] fluorescence correlation spectroscopy (FCS),[31,37] and near-field scanning optical microscopy.[40] Here, we studied how GM1 diffuses in DOPC SLBs by labeling it with 20 nm streptavidin-conjugated GNPs via biotinylated cholera toxin B subunits (CTxB), which bind to the oligosaccharide moiety of GM1 with high affinity.[41] We adjusted the average number of CTxB pentamers per GNP to be less than one (see Figure **1**B). The strong iSCAT signal made it possible to track the diffusion of single GNPs at 1.9 nm spatial precision and 1 kHz frame rate. The unlimited photostability of GNPs let us record as many as 5000 continuous steps in each trajectory, which was limited by our field of view in the current experiments.

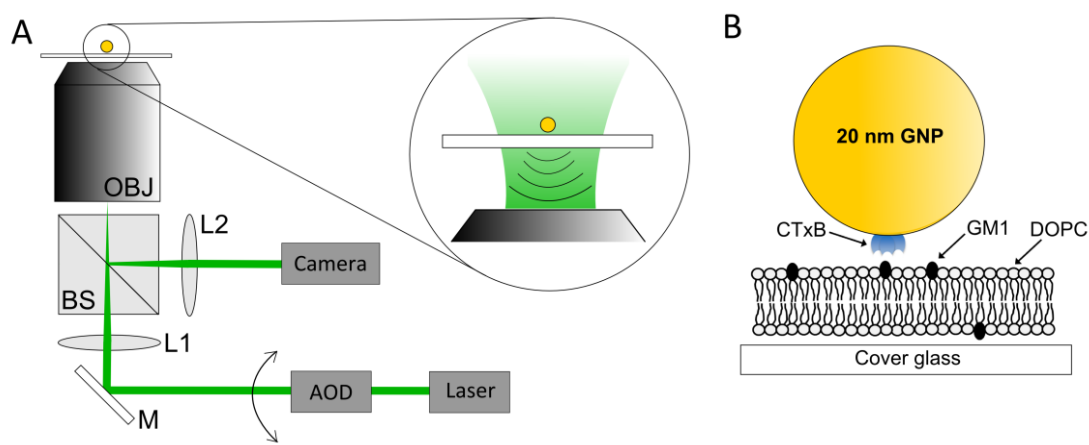

**Figure 1.** (A) Basic sketch of an iSCAT microscope. OBJ: objective, BS: 50/50 non-polarizing beam splitter, L1 and L2: lens, M: mirror. (B) Schematic illustration of a nanoparticle attaching to a GM1 molecule in lipid bilayers.

**RESULTS**

**Localization of 20 nm GNPs at 1.9 nm precision within 1 ms**

We recorded movies of the diffusion of 20 nm GNPs bound to GM1 at 1000 frames per second. A representative frame of a raw iSCAT video image of a single GNP is shown in Figure 2A. The scattering signal is superimposed by considerable systematic background contributions. The observed residual stripes are static and can be corrected by normalizing the raw data to a reference image taken before the measurement of the GNP. After applying this correction, the background only has slowly varying intensity (Figure 2B) caused by the inevitably non-uniform illumination of the sample. Since the particle of interest is mobile and diffusing across many pixels, the non-uniform illumination can be readily accounted for by finding the temporal median values of each pixel. After this correction, the processed image has no systematic noise and the dominant fluctuation stems from the shot noise of photons in the reference beam (see Figure 2C).

The dark spot in the image is the result of destructive interference between the scattered field and the reference beam. The position of the GNP is found by directly fitting this point-spread function with a two-dimensional Gaussian function. The camera has a full well capacity of 200k photoelectrons, leading to shot-noise background fluctuations of 0.22% for each pixel. Considering that the signal contrast of the 20 nm GNP is ~5%, this yields SNR~20. The theoretical limit of localization precision for this SNR is ~5% of the pixel size.[42] In our case, the optical magnification is chosen such that each pixel imaged a sample area of 38 × 38 nm² so that we would expect a precision of ~1.9 nm. Indeed, as shown by the histogram of localization errors in Figure 2D, we achieve this value experimentally, confirming that our detection is indeed shot-noise limited.

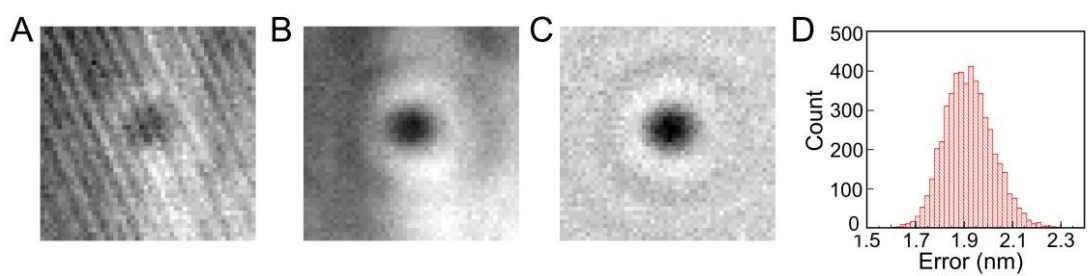

**Figure 2.** iSCAT image of a 20nm GNP on the model membrane recorded at 1000 Hz. (A) Raw data. (B) After correcting for the static stripes on the camera. (C) After correcting the non-uniform illumination. (D) Histogram of localization errors of a 20 nm GNP in consecutive 5000 frames of a video.

**Statistical analysis of diffusion trajectories**

Figure **3**A shows a representative diffusion trajectory (also see Movie S1, Supporting Material). The particle explores part of a ~2×2 μm² region within 5 seconds. The step size distributions in x and y directions at 1 ms time delay are plotted together in one histogram shown in Figure **3**B. There are 10,000 steps in the histogram as there are 5,000 steps in each lateral direction. The bin size of the histogram is chosen to be the localization precision of 1.9 nm. The high quality of the histogram allows us to see that the step size distribution is clearly not a Gaussian function. Instead, it can be described as a sum of two Gaussian functions corresponding to distinct diffusion coefficients of $D_1$ = 0.19 ± 0.01 μm²/s and $D_2$ = 0.028 ± 0.005 μm²/s with weighting factors $\varepsilon_1$ = 0.78 ± 0.04 and $\varepsilon_2$ = 0.22 ± 0.04, respectively. Here, $\varepsilon_1$ and $\varepsilon_2$ refer to the fractions of time that the particle undergoes diffusion at the corresponding mobilities.

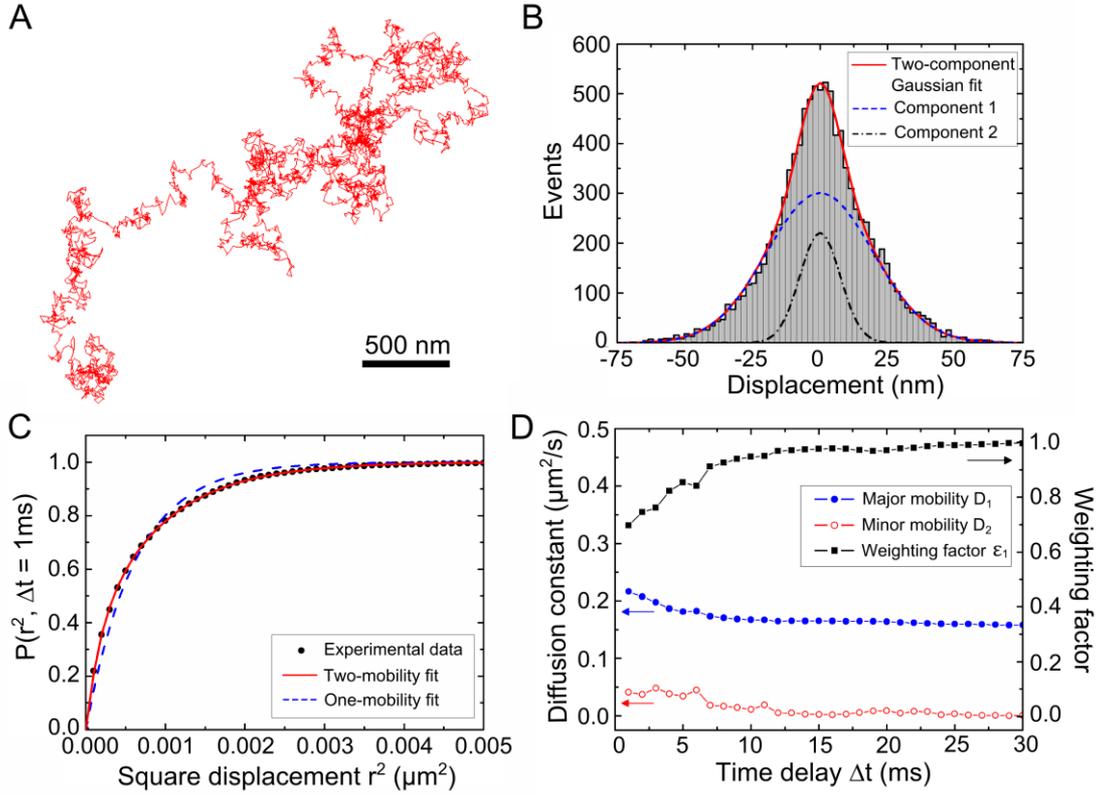

**Figure 3.** Representative results of an iSCAT SPT measurement on a single GNP bound to GM1 diffusing in a DOPC membrane with 1 mol% GM1. (A) GNP trajectory reconstructed from 5000 consecutive localizations at 1000 Hz. (B) Corresponding histogram of step sizes in both lateral directions (*bars*) along with a fit to a two-component Gaussian distribution (*red solid line*). Individual components are plotted as *blue dashed line* and *black dash-dotted line*, corresponding to diffusion constants of 0.19 µm²/s and 0.028 µm²/s with weighting factors of 0.78 and 0.22. (C) Corresponding cumulative probability distribution function of square displacements $P(r^2, \Delta t)$ for $\Delta t$ = 1 ms (•) along with fits to one-component (*blue dashed line*) and two-component (*red solid line*) models. The two-component fit results in major mobility $D_1$ = 0.22 µm²/s and minor mobility $D_2$ = 0.042 µm²/s with $\varepsilon_1$ = 0.69 and $\varepsilon_2$ = 0.31, respectively. (D) Dependence of major mobility $D_1$ (•), minor mobility $D_2$ (∘), and weighting factor $\varepsilon_1$ (■) on time delay $\Delta t$. For longer time delays, only a single effective mobility can be detected from the data.

An alternative analysis for finding multi-mobilities in one trajectory is to calculate the cumulative probability distribution of square displacements $P(r^2, \Delta t)$.[16] This quantity is equivalent to the probability of finding a particle within a circle of radius $r$ at time $\Delta t$ having started at the origin at time $t_0 = 0$. For the trajectory shown in Figure 3A, the measured $P(r^2, \Delta t = 1\text{ms})$ is plotted in Figure 3C. In this case, $P(r^2, \Delta t = 1\text{ms})$ can be fitted very well with a two-component distribution function

$$P(r^2, \Delta t) = 1 - \varepsilon_1 e^{-r^2/4D_1 \Delta t} - \varepsilon_2 e^{-r^2/4D_2 \Delta t}, \qquad (1)$$

where $D_1$ and $D_2$ represent the major and minor mobilities, respectively, and $\varepsilon_1$ and $\varepsilon_2$ are their corresponding weighting factors ($\varepsilon_1 + \varepsilon_2 = 1$). The biexponential fit yields $D_1 = 0.22 \pm 0.01$ µm²/s and $D_2 = 0.042 \pm 0.003$ µm²/s with $\varepsilon_1 = 0.69 \pm 0.02$ and $\varepsilon_2 = 0.31 \pm 0.02$. We point out that fitting the distribution with a one-component distribution function $P(r^2, \Delta t) = 1 - e^{-r^2/4D\Delta t}$ would increase the residual fitting errors by two orders of magnitude.

Next, we consider the probability distribution $P(r^2, \Delta t)$ of the trajectory in Figure 3A at increasing time delay $\Delta t$ from 1 ms to 30 ms. Each distribution is then fitted with a biexponential function (Eq. 1). In Figure 3D, it is shown how the resulting $D_1, D_2$, and $\varepsilon_1$ depend on the time delay $\Delta t$. We find that $\varepsilon_1$ approaches one as the time delay increases, indicating that diffusion at long delays is effectively reduced to a single mobility. Meanwhile, the major mobility $D_1$ decreases from 0.22 µm²/s to 0.16 µm²/s, and the minor mobility $D_2$ vanishes. The dependence of the diffusion constants on time delay shows that diffusion is clearly not normal on time scales below 10 ms.

We also analyzed the same trajectory by calculating the mean square displacement (MSD) for time delays $\Delta t$ ranging from 1 to 100 ms. Due to the high localization precision of our measurement, an accurate estimation of the microscopic diffusion coefficient $D_{\text{micro}} = 0.16$ µm²/s could be calculated from the first two data points in the MSD.[43] Furthermore, we fitted the MSD data with the function

$$\text{MSD}(\Delta t) = 4D_\alpha \Delta t^\alpha, \qquad (2)$$

where $D_\alpha$ is the anomalous diffusion coefficient and $\alpha$ is the anomalous diffusion exponent.[9] For the trajectory in Figure 3A, the resulting MSD is plotted in Figure 4. The anomalous diffusion coefficient $D_\alpha$ and the exponent $\alpha$ were found to be $0.18 \pm 0.004$ µm²/s and $1.02 \pm 0.006$, respectively, implying that the diffusion was close to normal. The fact that $\alpha$ turns out to be slightly greater than one is caused by the MSD noise at longer delay times. This example nicely demonstrates that the two-component diffusion behavior evident in the cumulative probability distribution is not easily captured by the MSD analysis (even when $\varepsilon_2$ is as large as 0.31).

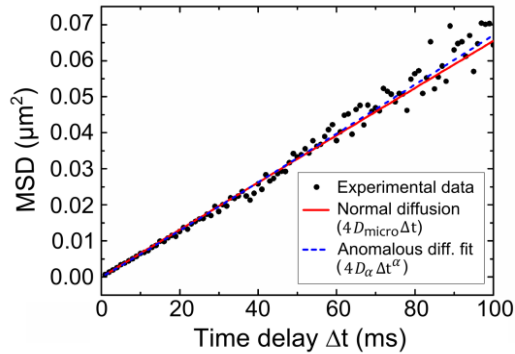

**Figure 4.** MSD calculated from the trajectory shown in Figure 3A (•). The *red solid line* is the extrapolation of the first two data points and thus represents a MSD for normal diffusion with $D_{\text{micro}} = 0.16$ µm²/s. The *blue dashed line* represents the fit using the anomalous diffusion model in Eq. 2 with $D_\alpha = 0.18$ µm²/s and $\alpha = 1.02$.

## Short-time nano-confinements

The high spatial and temporal resolutions of our system revealed occasional transient confinements in some of the recorded trajectories. Figure **5**A shows a trajectory (recorded from a DOPC membrane with 1 mol% GM1) that happens to have many such transient confinements. From the MSD analysis of this trajectory according to Eq. 2 (see Figure 5B), we obtain an anomalous diffusion exponent $\alpha$ = 1.03 ± 0.008 and coefficient $D_\alpha$ = 0.22 ± 0.006 µm²/s, suggesting that the diffusion is fairly slow and nearly normal. However, by carefully inspecting the trajectory data (see inset of Figure 5A), it becomes clear that the particle is transiently confined within regions of extension ~20 nm for tens of milliseconds. Indeed, these transient confinements over short time intervals are not apparent in the MSD analysis because they are averaged out in the analysis.

The trajectories that display substantial amount of confinements can be, nevertheless, modeled when considering their $P(r^2, \Delta t)$ for $\Delta t$ = 1ms if one allows for a third mobility component (Figure **5**C):

$$P(r^2, \Delta t) = 1 - \varepsilon_1 e^{-r^2/4D_1\Delta t} - \varepsilon_2 e^{-r^2/4D_2\Delta t} - \varepsilon_3 e^{-r^2/4D_3\Delta t}. \quad (3)$$

Including the third component reduces the residual fitting error by more than one order of magnitude. Fitting with four components is not considered since it reduces the error by less than one order of magnitude, and it turns out not to be stable. For comparison, the fits according to single-, two-, and three-component models are plotted in Figure **5**C. The results from the three-component fit are $D_1$ = 0.009 ± 0.001 µm²/s, $D_2$ = 0.11 ± 0.01 µm²/s, $D_3$ = 0.75 ± 0.04 µm²/s with $\varepsilon_1$ = 0.43 ± 0.01, $\varepsilon_2$ = 0.39 ± 0.01, $\varepsilon_3$ = 0.18 ± 0.01 (note that the order of these three mobilities are determined by their weighting factors, not by their diffusion coefficients). The large fraction of $\varepsilon_1$ for $D_1$ shows that it is necessary to include a small diffusion associated with the strong local confinements. About ~2% of the trajectories in DOPC membranes with 1 mol% GM1 show the third mobility.

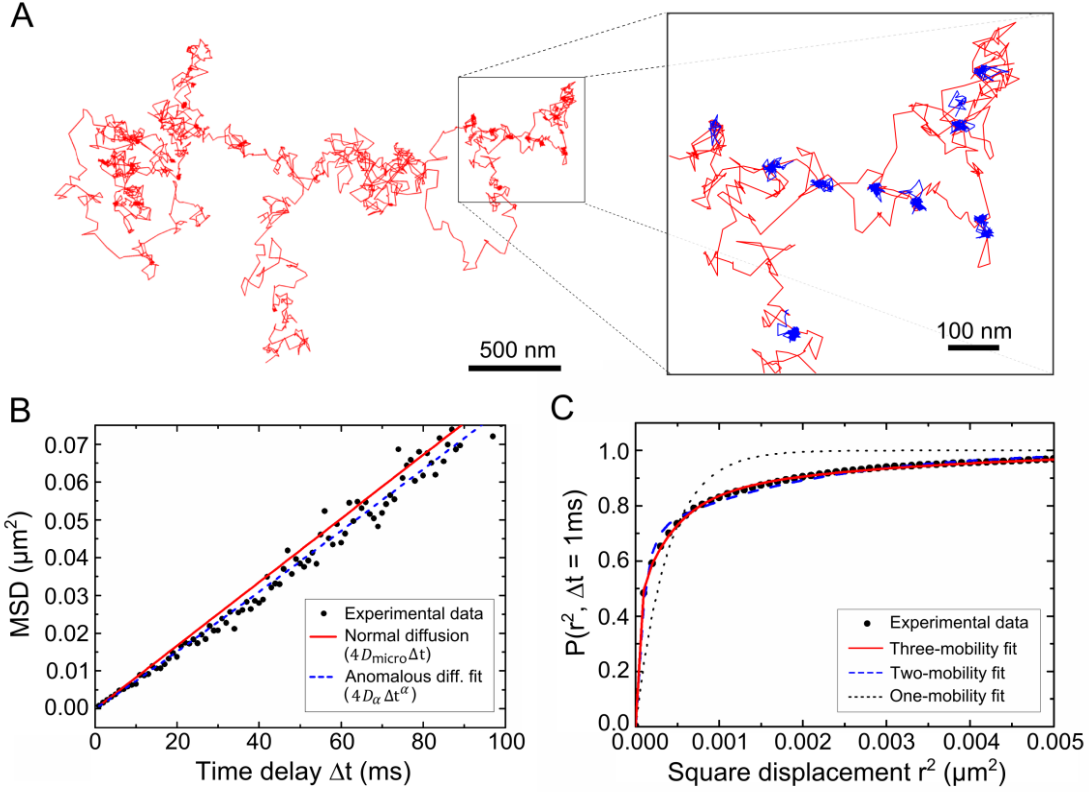

**Figure 5.** (A) Trajectory of a single GNP bound to GM1 in a DOPC SLB with 1 mol% GM1. The particle undergoes numerous transient confinements within regions of 20 to 40 nm in size for some tens of milliseconds. Using the approach described in the text, we reliably detect these confinements (blue parts of the trajectory in the *inset*). (B) Corresponding MSD analysis of the trajectory in (A) with resulting microscopic diffusion coefficient $D_{\text{micro}}$ = 0.21 μm²/s. The *red solid line* represents the extrapolation of the first two data points (i.e., MSD = $4D_{\text{micro}}\Delta t$). The *blue dashed line* represents the fit using Eq. 2 with $D_\alpha$ = 0.22 μm²/s and $\alpha$ = 1.03, indicating the diffusion is nearly normal. (C) Corresponding $P(r^2, \Delta t)$ of the trajectory for time delay $\Delta t$ = 1ms (•) along with fits to the one-mobility (*black dotted*), two-mobility (*blue dashed*), and three-mobility (*red solid*) models. The three-mobility model fits the data best with $D_1$ = 0.009 μm²/s, $D_2$ = 0.11 μm²/s, $D_3$ = 0.75 μm²/s and corresponding $\varepsilon_1$ = 0.43, $\varepsilon_2$ = 0.39, and $\varepsilon_3$ = 0.18, respectively.

The presence of transient confinements suggests that the membrane was highly heterogeneous. Interestingly, these confinement domains are frequently revisited, indicating that the traps are stationary within our observation window of the order of seconds. In order to analyze the confinement sizes and times, the short-time mobility of the particles was calculated. In particular, a transient MSD was calculated as a function of time for a sliding time window of length $T$ and a single time delay $\Delta t$ as follows:

$$\text{MSD}_{\text{trans}}(t) = \langle \text{MSD}(\Delta t) \rangle_T(t) = \frac{1}{T - \Delta t} \sum_{\tau=t}^{t+T-\Delta t} \left( \mathbf{r}(\tau) - \mathbf{r}(\tau + \Delta t) \right)^2. \quad (4)$$

Here, $\mathbf{r}(\tau)$ is the position of the particle at time $\tau$. A drop of $\mathrm{MSD}_{\mathrm{trans}}(t)$ below a certain threshold $\mathrm{MSD}_{\mathrm{thresh}}$ was counted as a confinement. For the data presented here we used $\mathrm{MSD}_{\mathrm{thresh}} = (47\,\mathrm{nm})^2$. As shown in the inset in Figure 5A, using this approach one can reliably detect the nanoscopic confinements in the trajectories. The results of the analysis of the confinement times and sizes are shown in Figure 6. As can clearly be seen from these data, the majority of GNPs are confined within radii of 17 to 35 nm for less than 20 ms, and there is no dependence of confinement times and sizes on GM1 concentration.

To examine this intriguing phenomenon further, we replaced the GM1 lipids by DOPE lipids, which are more compatible with the host DOPC membrane in terms of structure and size. Here, we used head-group-biotinylated DOPE (DOPE) and streptavidin-conjugated 20 nm GNPs as label. In contrast to the above-mentioned GM1 measurements, where GNPs had only one CTxB linker with maximal five binding sites, the GNPs in this experiment had a large coverage of streptavidin. Interestingly, we again detected transient nanoscale confinements in a number of trajectories with the same distributions of confinement times and sizes as for GM1 (see Figure 6). The fact that the confinement time and size distributions depend neither on concentration nor on the type of the lipid probe are consistent with the hypothesis that the observed confinement events are caused by substrate effects. We emphasize that these short-time and nanoscopic confinements would not have been detectable with low temporal or spatial resolutions. Future experiments that combine AFM and iSCAT could search for correlations between the nano-confinements and substrate/membrane topography.

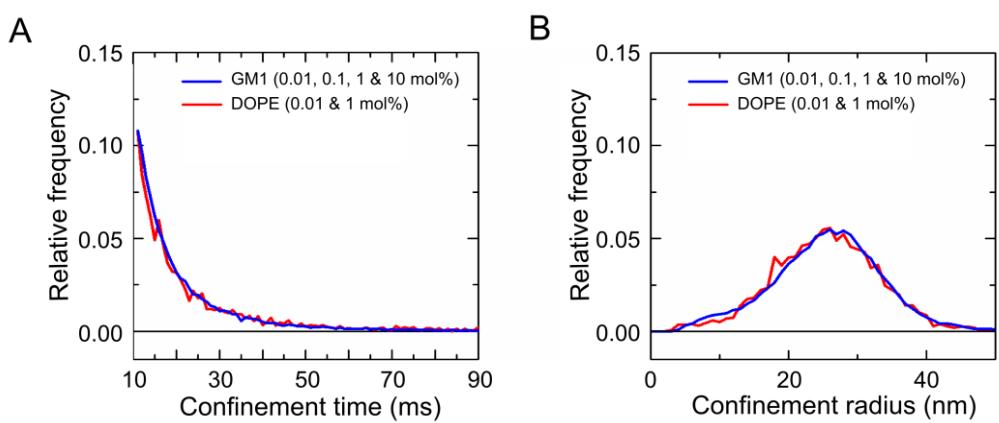

**Figure 6.** Distributions of times (A) and radii (B) of nanoscopic confinements of GNPs bound to GM1 and DOPE diffusing in DOPC membranes.

**The effect of concentration on the diffusion behavior**

To investigate the effect of the concentration, we recorded diffusion trajectories over four decades of GM1 concentrations at 0.01, 0.1, 1, and 10 mol%. In order to have sufficient statistical data for each trajectory, only those that were longer than 1000 steps were analyzed. 557, 800, 640, and 828 trajectories were collected from samples at concentrations of 0.01, 0.1, 1, and 10 mol%, respectively. We then calculated $P(r^2, \Delta t = 1 \text{ ms})$ for each trajectory and fitted the distribution by both two-component and three-component models as described in Eq. 1 and Eq. 3. Fitting with the three-component model always provides a lower residual error, but we only considered the three-component model when the error was reduced by at least one order of magnitude. Among the trajectories that were fitted by the two-mobility model, we classified a trajectory as normal diffusion if the weighting factor of the major mobility was greater than 0.95. Following these criteria, the trajectories were categorized into one-, two-, or three-mobility diffusion, and their fractions measured for different GM1 concentrations are plotted in Figure **7**. We find that the percentage of one-mobility diffusion decreases from 25.7% to 7.4% when the GM1 concentration increases from 0.01 mol% to 10 mol%. The results clearly indicate that more GM1 molecules deviate from normal diffusion at higher concentrations. The fraction of the three-mobility diffusion is low (ranging from 0.1% to 2.8%) at all GM1 concentrations.

We also recorded diffusion trajectories of DOPE lipids in DOPC SLBs at 0.01 mol% (331 trajectories) and 1 mol% (249 trajectories) concentrations. The resulting fractions of one-, two-, and three-mobility diffusion are shown in Figure 7. As compared to our findings for GM1, DOPE exhibits larger fractions of one-mobility diffusion (> 25%), implying freer diffusion.

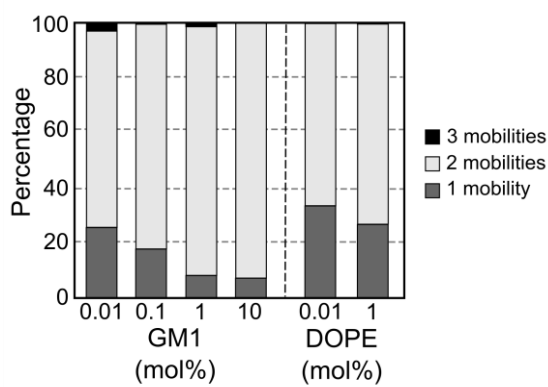

**Figure 7.** Fractions of one-, two-, and three-mobility diffusion of GNPs bound to GM1 and DOPE observed in DOPC SLBs at various concentrations.

Except for the small portion of trajectories that undergo three-mobility diffusion, the behavior of the trajectories can be portrayed in the histograms of the major and minor diffusion constants $D_1$ and $D_2$, and the weighting factors $\varepsilon_1$ (see Figure **8**). The results of GM1 at four concentrations are plotted in the top four rows of Figure 8. The histograms of $D_1$ show two distinct populations (see left plots of Figure **8**): one main population centered at ~0.2 µm²/s and a smaller one at a larger mobility of 0.9 to 1.2 μm²/s. The major peak centered at 0.2 μm²/s does not change substantially over four orders of magnitude of GM1 concentration. On the other hand, the mean diffusion coefficient of the small population at larger mobilities shifts towards a lower value (from 1.2 μm²/s to 0.9 μm²/s) for higher GM1 concentration. In fact, those trajectories that display a large $D_1$ usually display normal diffusion (see the relationship between $D_1$ and $\varepsilon_1$ in Figure **S1** in the Supporting Material). Meanwhile, the histograms of $D_2$ (middle figures of Figure **8**) show that almost all particles have a very low minor mobility without a noticeable dependence on GM1 concentration. The histograms of $\varepsilon_1$ (right figures of Figure **8**) show a broad distribution from 0.5 to 1 along with a narrow peak at one. The latter represents the trajectories that undergo one-mobility diffusion. It can be seen that the peak at $\varepsilon_1 \sim 1$ diminishes when GM1 concentration increases, meaning that the probability of one-mobility diffusion decreases at higher GM1 concentrations. Scatter plots of $D_1 - D_2$, $D_1 - \varepsilon_1$, and $D_2 - \varepsilon_2$ measured at these four GM1 concentrations are shown in Figure **S1** in the Supporting Material. We remark that a standard MSD analysis would not have detected the concentration-dependent behavior presented in this section.

In Figure 8, we also present the histograms of $D_1$, $D_2$ and $\varepsilon_1$ of the control experiments on DOPE (see bottom two rows). No considerable difference is observed between the two concentrations of DOPE. However, instead of two populations with narrow distributions as was the case for GM1, the histogram of $D_1$ for DOPE shows a broad distribution with average values of 0.56 μm²/s and 0.81 μm²/s for concentrations of 0.01 mol% and 1 mol%, respectively. Moreover, the $\varepsilon_1$ histogram of DOPE also yields a higher frequency of large $\varepsilon_1$ values, which is consistent with having a larger fraction of one-mobility diffusion (see Figure 7).

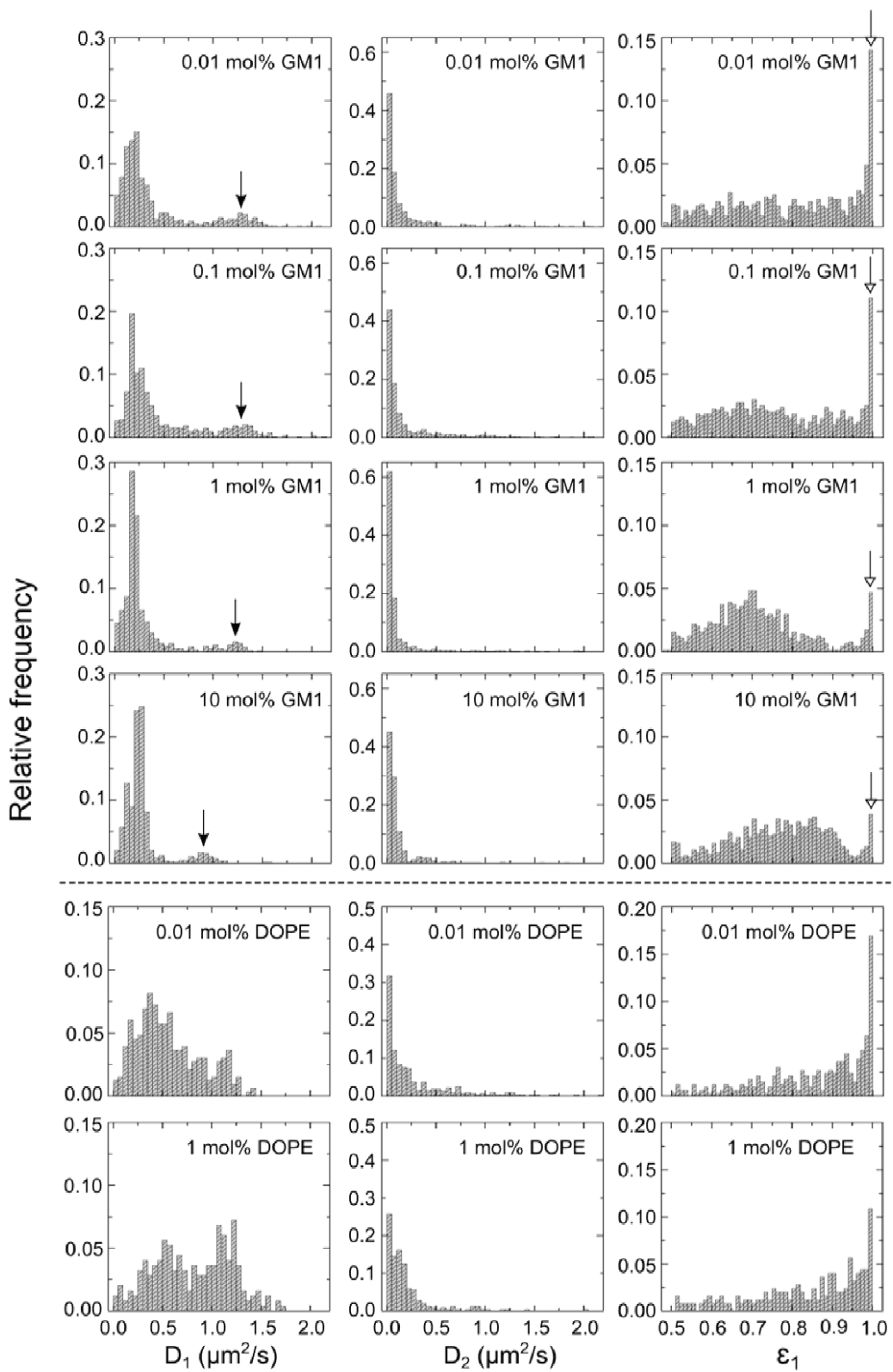

**Figure 8.** Results of the cumulative probability distribution analysis for GM1 and DOPE diffusion in DOPC membranes with four GM1 concentrations: (*first row*) 0.01 mol%, (*second row*) 0.1 mol%, (*third row*) 1 mol%, (*fourth row*) 10 mol%, and two DOPE concentrations: (fifth row) 0.01 mol%, (sixth row) 1 mol%. *Left*: histograms of the major mobility $D_1$; *Middle*: histograms of the minor mobility $D_2$, and *Right*: histograms of weighting factor $\varepsilon_1$. The *solid arrows* in the left plots of GM1 indicate that the small

population of larger mobilities of $D_1$ shifts towards lower values as the GM1 concentration increases. The *hollow arrows* in the right plots of GM1 mark the decreasing frequency of $\varepsilon_1 \sim 1$ as the GM1 concentration increases, implying that more GM1 lipids deviate from normal diffusion in DOPC SLBs with higher GM1 concentration. Results of DOPE show faster diffusion with larger value of $D_1$ and freer diffusion with larger frequency of $\varepsilon_1 \sim 1$. Diffusion of DOPE at 0.01 mol% and 1 mol% does not display noticeable differences.

**DISCUSSION**

The combination of iSCAT SPT with a spatial precision of <2 nm and temporal resolution of 1 ms together with a long observation time allows us to carefully investigate the diffusion of lipid molecules in membranes. We observe that GM1 undergoes diffusion with multiple mobilities in DOPC SLBs. This behavior has also been detected using fluorescence-based SPT and FCS,[16,37] but unlike the previous studies where the data were collected either from ensemble averages or multiple short trajectories, we clearly find multiple mobilities in individual trajectories and resolve intriguing nanoscopic confinements.

We find that most GM1-CTxB-GNP complexes diffuse at $D_1 \sim 0.2$ μm²/s. This slow diffusion is in agreement with previous observations,[36,37] which have discussed different underlying mechanisms. Burns *et al.* studied diffusion of GM1 in the DOPC-rich phase of phase-separated DOPC/DPPC SLBs by FCS and observed that the GM1 diffusion coefficient decreases by at least six times upon binding of CTxB.[37] They attributed this to an enhanced interleaflet coupling of CTxB-induced clusters of GM1. Here, one has to bear in mind that CTxB has a pentameric structure and thus may bind up to five GM1 molecules in a cluster. Furthermore, the larger size of the CTxB-GM1 complexes could also cause slower mobility either directly[44] or via stronger substrate interaction.[45,46]

In our control experiments, we chose to measure the diffusion of head-group-biotinylated DOPE because of its unsaturated alky chains and minimal tendency to form clusters. The distinct differences in the mobilities of GM1 and DOPE at various concentrations (see Figures 7 and 8) indicate that DOPE undergoes freer and faster diffusion. This observation indicates that the slow diffusion of GM1 is largely the result of GM1 clustering.

Recent works have recognized that lipid mobility in SLBs can be influenced by the supporting substrate.[7,47-49] For example, SLBs were observed to follow the atomic structure of a substrate,[50] and single atomic steps on the supporting substrate were found to cause anomalous diffusion.[14] In our case, the surface roughness of the high-quality borosilicate glass had height variations of the order of nanometer (typical root mean-square roughness of 0.3 to 0.8 nm) over lateral extensions of tens of nanometers (measured by AFM; data not shown). We can also deduce membrane heterogeneities from the iSCAT measurements by estimating the lateral spatial scale of diffusion as $\sqrt{\mathrm{MSD}(\Delta t)}$ with effective short-time diffusion coefficient $D_{\mathrm{micro}}$. Thus, if we use the crossover from the multi-mobility diffusion to normal diffusion (Figure 3D) at time interval ~10 ms as an indicator for the onset of heterogeneity effects, we arrive at a lateral scale ranging from 25 to 80 nm over a period of $\Delta t$ = 1 to 10 ms. To this end, our findings seem consistent with the previous observations.

Comparing the diffusion behavior of fast and slow diffusing particles in our measurements, it can be seen that fast diffusing particles tend to diffuse more often with one mobility ($\varepsilon_1$ is close to one) than the slow particles. This suggests that smaller GM1 clusters not only diffuse faster, but their diffusion is also nearly normal, possibly because less substrate interaction is experienced due to their smaller sizes. Another noticeable trend is that the percentage of normally diffusing particles ($\varepsilon_1 \sim 1$) decreases as the GM1 concentration increases. This could be explained by more GM1 molecules diffusing at slower rates and with multiple mobilities when the GM1 concentration increases, leading to more GM1 clusters of large size.

In the present work, we used GNPs with a diameter of 20 nm, which is considerably smaller than the size of the particles used in many previous studies.[17-20] Using small labels minimizes the load of the tracked lipids and most importantly makes monovalent binding more accessible. Multivalent binding can lead to crosslinking of lipids in a membrane, and therefore not only influence their diffusion constants, but also possibly initiate large-scale phase separation.[51] As an additional measure to avoid multivalent binding, we furthermore conjugated fewer than one CTxB pentamer per GNP on average. We remark in passing that the consistency between previous fluorescence-based SPT works and our current study shows that labeling with 20 nm GNPs does not perturb lipid diffusion.

The temporal resolution of iSCAT microscopy in our experiments was limited by the acquisition rate of the camera. Considering that new CMOS cameras are able to record videos at a MHz frame rate, we believe iSCAT holds a great promise for reaching microsecond time resolution. Here, one has to adjust the excitation intensity to obtain a sufficiently large scattering signal that allows short-time integration and, thus, higher recording speed. Our preliminary experiments indicate that this is indeed feasible.

**CONCLUSION AND OUTLOOK**

We demonstrated localization of single 20 nm GNPs at nanometer precision within one millisecond, using moderate excitation intensity of < 10 kW/cm$^2$. Long trajectories with a large number of steps allowed us to accurately identify multiple mobilities displayed in single diffusion trajectories without the need for ensemble averaging. Furthermore, we found strong transient local confinements within areas as small as 20 nm. The observed deviations from normal diffusion point to membrane heterogeneities, which are most likely induced by the underlying substrate. These results show the power of high-speed iSCAT SPT for investigating complex diffusion phenomena such as compartmentalization and lipid rafts in cell membranes.[1-3]

The extension of our measurements from model membranes to plasma membranes of living cells will be an exciting next step that would offer unprecedented spatial and temporal resolutions. The challenge of such an effort would be in distinguishing the signal of the particle to be traced from the background scattering of the constituents of biological cells. Two measures can address this issue. First, the plasmon resonance of metallic nanoparticles offers a convenient signature to discriminate the particle response against the scattering of the cell.[27] Second, differential measurements between consecutive frames can help identify the moving parts of the image.

A fascinating prospect of iSCAT is three-dimensional tracking of nanoparticles. Indeed, previous works have shown that the axial position of a particle can be extracted from the contrast of the point-spread function, which in turn encodes the phase of the detected scatter field.[52] Finally, iSCAT lends itself to combination with AFM measurements, which could identify correlations between local nanoscopic confinements and topographical information.

**ACKNOWLEDGMENTS**

This project was financed by the Alexander von Humboldt Professorship and the Max Planck Society. We thank Philipp Kukura and Ana-Sunčana Smith for stimulating discussions and Cornelia Becker for technical support.

## MATERIALS AND METHODS

### Materials

1,2-dioleoyl-sn-glycero-3-phosphocholine (DOPC), brain monosialotetrahexosylganglioside (GM1), and 1,2-dioleoyl-sn-glycero-3-phosphoethanolamine-N-cap-biotinyl (DOPE) were purchased from Avanti Polar Lipids (Alabaster, AL). Dye-labeled lipid atto532-1,2-dioleoyl-sn-glycero-3-phosphoethanolamine (atto532-DOPE) was purchased from ATTO-TEC GmbH (Siegen, Germany). Lipids were stored in chloroform at a concentration of 10 mg/ml at -20° C. Streptavidin-conjugated 20 nm GNPs were purchased from BBI (Cardiff, UK). Biotinylated cholera toxin B subunit (CTxB) was purchased from Life Technologies (Frankfurt, Germany). Hellmanex III was purchased from Hellma Analytics (Müllheim, Germany). Phosphate buffered saline (PBS) and HEPES buffer were purchased from Sigma-Aldrich (Taufkirchen, Germany). All chemicals and solvents were of analytical grade and used without further purification. Milli-Q water (Merck Millipore, Billerica, MA) was used to prepare all solutions.

### Membrane preparation

SLBs were prepared by vesicle fusion on a glass substrate.[53] Lipid mixtures containing DOPC and GM1 or DOPE were freshly prepared before membrane formation. A trace amount (<0.1 mol%) of atto532-DOPE was added for fluorescence observation. The lipid mixture was dried in a glass vial with a stream of nitrogen and put in an exsiccator overnight to completely remove the solvent. The dried lipids were hydrated in buffer solution (150 mM NaCl and 10 mM HEPES) for at least one hour (final lipid concentration: 1 mg/ml). Multilamellar vesicles (MLVs) were then formed by vortexing the lipid solution. Next, the MLV solution was tip-sonicated (Q700, Qsonica, Newtown, CT) in a water-ice bath to break the MLVs into small unilamellar vesicles (SUVs) with a power of 175 W in pulse mode (50% duty cycle for 1 sec) for around 20 minutes or until the solution is clear. The lipid solution was centrifuged for 20 min at 16000g and 4 °C to remove titanium particles from the sonicator tip as well as big residual vesicles from the SUV suspension. The supernatant containing SUVs was collected and diluted 4 times in buffer solution to a final lipid concentration of less than 250 μg/ml. The SLBs were prepared on a chambered cover glass (8-well, LabTek II, Thermo Scientific, Waltham, MA), which was pre-cleaned by incubating in 2% Hellmanex overnight followed by bath sonication in 2% Hellmanex, 1M KOH, and deionized water sequentially for 15 minutes each. The glass was dried and then treated by oxygen-plasma for 10 minutes before use. For membrane preparation, each chamber was filled with 200 μl of the SUV suspension. Next, $CaCl_2$ was added to each chamber to the concentration of 2 mM to enhance vesicle bursting. After 30 to 60 minutes of incubation time, the membranes were washed rigorously with the buffer solution to remove excess SUVs.

**Conjugation of GNPs with CTxB**

CTxB binds to the oligosaccharide moiety of the GM1 molecules with high affinity [41] and was therefore used as a cross-linker for specifically binding GNPs to GM1 in the membrane (see Figure **1**B). Streptavidin-conjugated GNPs were functionalized with biotinylated CTxB. The conjugation was carried out at room temperature for 3 hours and the average number of CTxB pentamers per GNP was controlled to be less than one. After the incubation, unconjugated CTxB was removed by centrifugation at 8000g and 4 °C for 20 minutes and dumping the supernatant. The CTxB-GNP conjugates were re-suspended in deionized water.

**GM1 and DOPE labeling in SLBs by conjugated GNPs**

The CTxB-GNP conjugates or streptavidin conjugated GNPs were added to the freshly prepared DOPC/GM1 or DOPC/DOPE membrane, respectively, under continuous monitoring with iSCAT microscopy. Individual attachment events can be clearly visualized by iSCAT when GNPs approach the membrane and stay in focus during diffusion. After a few minutes of incubation, the unattached particles in the solution were removed by gentle buffer exchange. To assure that the CTxB-GNPs bind specifically to GM1 in the membrane, two control experiments were performed. First, GNPs without CTxB were added to membranes containing GM1. Second, CTxB-GNPs were added to membranes without GM1. In both cases, no diffusing GNPs were observed in the membrane. All measurements were performed at 22 °C.

# SUPPORTING MATERIAL

Movie S1. The iSCAT video shows a single GNP bound to GM1 diffusing in a DOPC membrane with 1 mol% GM1. The raw data was recorded at 1000 frames per second, but here it was down-sampled and played at 25 frames per second for visualization. The resolution of this movie was also increased to $500 \times 500$ without interpolation for better visualization.

Figure S1 displays scatter plots of $D_1 - D_2$, $D_1 - \varepsilon_1$ and $D_2 - \varepsilon_2$ measured from DOPC SLBs with various GM1 concentrations (from 0.01% to 10 mol%). In the left scatter plots, $D_1$ and $D_2$ show clear positive correlations. The two clusters mirrored in the line with a slope of one result from the exchanging roles of the major and minor mobilities (notice that the definition of $D_1$ and $D_2$ are major and minor mobilities, not fast and slow mobility). In the middle scatter plots, it can be seen that the majority of the particles diffuse at $D_1$ ~0.2 µm²/s with $\varepsilon_1$ ranging from 0.5 to 1. It is worth noting that for those trajectories with $\varepsilon_1$ close to one $D_1$ is usually large (>0.5 µm²/s, indicated by red ellipses). This correlation suggests that fast diffusing particles usually diffuse more normally. In the right scatter plots, the minor mobility $D_2$ shows weak dependence on the weighting factor $\varepsilon_2$.

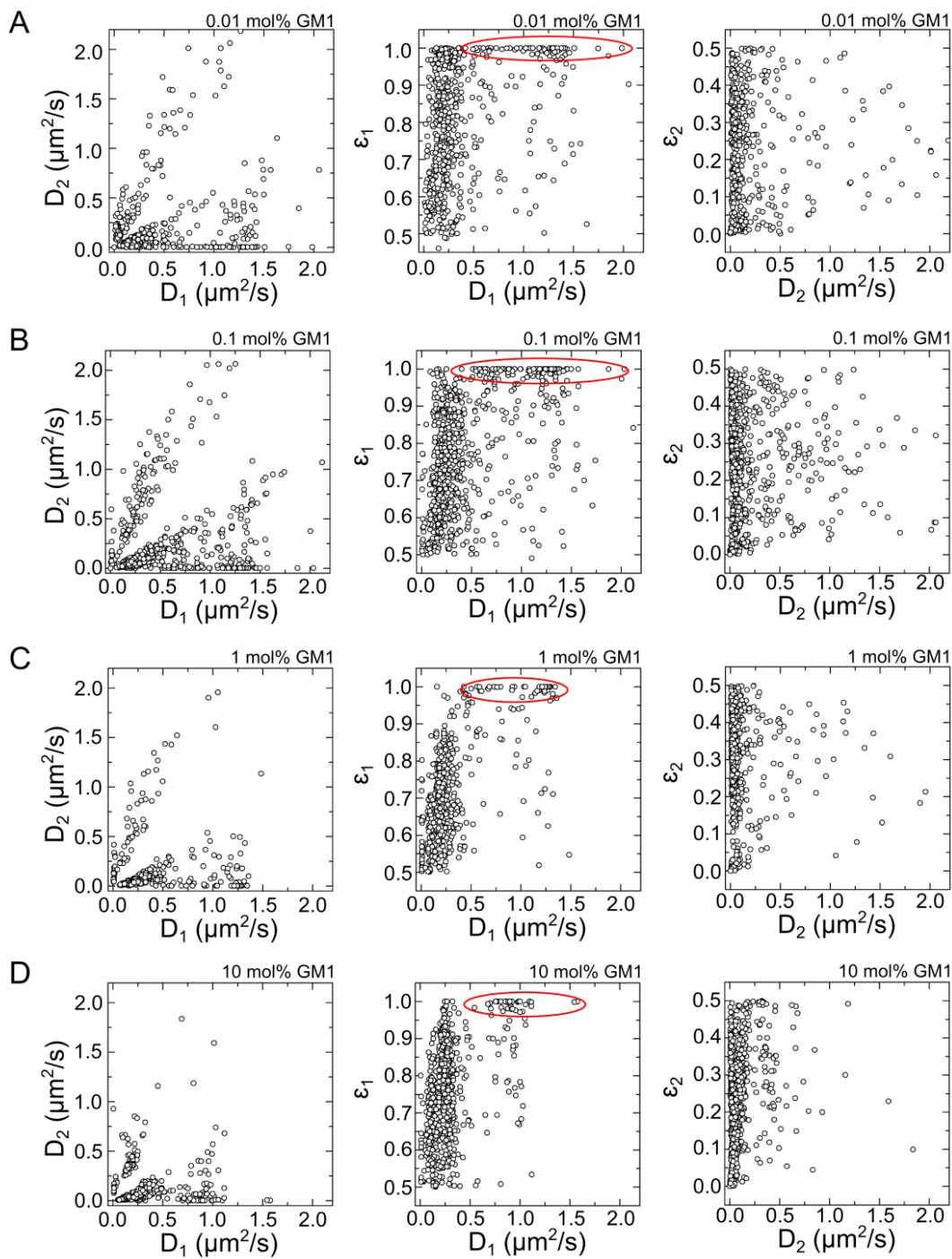

**Figure S1**. Scatter plots of $D_1 - D_2$ (*left*), $D_1 - \varepsilon_1$ (*middle*) and $D_2 - \varepsilon_2$ (*right*) measured from DOPC SLBs with GM1 concentrations of 0.01 mol% (A), 0.1 mol% (B), 1 mol% (C), and 10 mol% (D). The red ellipses indicate the particles that undergo fast and normal diffusion.